\documentstyle[prc,aps,epsf]{revtex}

\begin{document}
\title{Correlation effects in single--particle overlap functions and
one--nucleon removal reactions}
\author{M.K. Gaidarov, K.A. Pavlova, S.S. Dimitrova, M.V. Stoitsov, and
A.N. Antonov}
\address{Institute of Nuclear Research and Nuclear Energy, Bulgarian\\
Academy of Sciences, Sofia 1784, Bulgaria}
\author{D. Van Neck}
\address{Laboratory for Theoretical Physics, Proeftuinstraat 86, B-9000 Gent,\\
Belgium}
\author{H. M\"{u}ther}
\address{Institut f\"{u}r Theoretische Physik, Universit\"{a}t T\"{u}bingen,\\
Auf der Morgenstelle 14, D-72076 T\"{u}bingen, Germany}
\maketitle

\begin{abstract}
Single--particle overlap functions and spectroscopic factors are calculated
on the basis of the one--body density matrices (ODM) obtained for the nucleus
$^{16}O$ employing different approaches to account for the effects of
correlations. The calculations use the relationship between the overlap functions
related to bound states of the $(A-1)$--particle system and the ODM for the ground
state of the $A$--particle system. The resulting bound--state overlap functions are
compared and tested in the description of the experimental data from $(p,d)$
reactions for which the shape of the overlap function is important.
\end{abstract}

\vspace{1cm}

\section{Introduction}

The strong short-range and tensor components of the nucleon--nucleon
interactions induce correlations in the nuclear wave function which are
going beyond the independent--particle approximation, e.g., the
Hartree--Fock method. Therefore it has always been a point of experimental
and theoretical interest to find observables, which reflect these
correlations in a unambiguous way. In this sense both, the overlap functions
and single--nucleon spectroscopic factors, have attracted much attention in
analyzing the empirical data from one--nucleon removal reactions, such as $%
(e,e^{\prime }p)$, $(p,d)$, $(d,^{3}He)$ , see e.g. \cite
{An93,Lap93,Bo94,Bl95} and also in other domains of many--body physics, as
e.g. atomic and molecular physics \cite
{Vn93,An95,Be65,Vn96,Vn97,Vn98,Sto96,Ge96}.

Recently, a general procedure has been adopted \cite{Vn93} to extract the
bound-state overlap functions and the associated spectroscopic factors and
separation energies on the base of the ground-state one-body density matrix.
The advantage of the procedure is that it avoids the complicated task for
calculating the whole spectral function in nuclei. One is able instead to
incorporate the knowledge of realistic one-body density matrices emerging
from various correlation methods going beyond the independent--particle
picture which have been proposed over the years \cite
{An93,An88,Ma91,Pi92,Co92,De83,Ja55,Sa96,Fa98,Sa97,Am98,Di92,Po96,Mu95,Mut95,Ci96,Mu94}%
.

Initially, the procedure for extracting bound-state overlap functions has
been applied \cite{Sto96} to a model one-body density matrix \cite{Sto93}
accounting for the short--range nucleon correlations within the low--order
approximation to the Jastrow correlation method (JCM). The calculations were
based on a single harmonic--oscillator Slater determinant and Gaussian--like
state--independent correlation functions. The resulting overlap functions have
been used \cite{Di97} to study one--nucleon removal processes in contrast to
the mean--field approaches which account for the nucleon correlations by
modifying the mean--field potentials. The results obtained for the
differential cross--sections of $^{16}O(p,d)$ and $^{40}Ca(p,d)$ pick--up
reactions at various incident energies demonstrated that the overlap functions
can be applied as realistic form factors to evaluate absolute cross--sections of such
reactions. Of course, the general success of the above procedure depends
strongly on the availability of realistic one-body density matrices.

This work can be considered as an extension of the analysis of
single--particle overlap functions based on the procedure \cite{Vn93} to
more realistic one--body density matrices emerging from the correlated basis
function (CBF) method \cite{Vn97,Sa96,Sa97} and the Green function method
(GFM) \cite{Po96}. We have chosen the CBF theory, based on the Jastrow approach
\cite{Ja55}, since it is particularly suitable for the study of the
short--range correlations in nuclei. So far the calculations have been
performed for infinite nuclear matter and some light nuclei as e.g. the
variational Monte Carlo calculations for the $^{16}O$ nucleus \cite{Pi92}.
The CBF calculations have recently been extended to medium--heavy
doubly-closed shell nuclei \cite{Co92,Sa96,Sa97,Am98} using various levels of the
Fermi hypernetted chain (FHNC) approximation \cite{Sa96,Fa98}. The Green
function method \cite{Di92,Po96} provides detailed information on the
spectral functions and nucleon momentum distributions \cite{Mu95,Mut95}
predicting the largest effects of the short--range and tensor correlations
at high momentum and energy \cite{Ci96,Mu94}.

The main purpose of this work is twofold. Using the procedure \cite{Vn93},
we first calculate all bound-state overlap functions on the basis of
one--body density matrices emerging from the CBF and Green function
methods for the $^{16}O$ nucleus in order to analyze and compare their
properties in coordinate and momentum spaces. Then, the resulting overlap
functions are tested in the description of the experimental data from $%
^{16}O(p,d)^{15}O$ reaction for which the shape of the overlap functions is
important. Such an investigation allows to examine the relationship
between the one-body density matrix and the associated overlap functions
within the correlation methods used and also to clarify the importance of
the effects of NN correlations on the overlap functions and $(p,d)$
cross--sections.

Some basic relations of the methods used to determine the effects of
correlations on the one--body density matrices are given in Section
II. In Section III we present numerical results for the quantities under
consideration in the case of $^{16}O$. Section IV contains a summary and
conclusions.

\section{Correlated nuclear wave functions}

The single--particle overlap functions in quantum--mechanical many--body
systems are defined by the overlap integrals between eigenstates of the $A$%
--particle and the $(A-1)$--particle systems:
\begin{equation}
\phi _{\alpha }({\bf r})=\langle \Psi _{\alpha }^{(A-1)}|a({\bf r})|\Psi
^{(A)}\rangle ,  \label{eq:1}
\end{equation}
where $a({\bf r})$ is the annihilation operator for a nucleon with spatial
coordinate ${\bf r}$ (spin and isospin operators are implied). In the
mean--field approximation $\Psi ^{(A)}$ and $\Psi _{\alpha }^{(A-1)}$ are
single Slater determinants and the overlap functions are identical with the
mean--field single--particle wave functions. Of course, this is not the case
at the presence of correlations where both, $\Psi ^{(A)}$ and $\Psi _{\alpha
}^{(A-1)}$, are complicated superpositions of Slater determinants. In
general, the overlap functions (1) are not orthogonal. Their norm defines
the spectroscopic factor
\begin{equation}
S_{\alpha }=\langle \phi _{\alpha }|\phi _{\alpha }\rangle .  \label{eq:2}
\end{equation}
The normalized overlap function associated with the state $\alpha $ then
reads
\begin{equation}
\tilde{\phi}_{\alpha }({\bf r})=S_{\alpha }^{-1/2}\phi _{\alpha }({\bf r}).
\label{eq:3}
\end{equation}
The one--body density matrix can be expressed in terms of the
overlap functions in the form:
\begin{equation}
\rho ({\bf r},{\bf r^{\prime }})=\sum_{\alpha }\phi _{\alpha }^{*}({\bf r}%
)\phi _{\alpha }({\bf r^{\prime }})=\sum_{\alpha }S_{\alpha }\tilde{\phi}%
_{\alpha }^{*}({\bf r})\tilde{\phi}_{\alpha }({\bf r^{\prime }}).
\label{eq:4}
\end{equation}

It has been shown in \cite{Vn93} that the one-body overlap functions (\ref
{eq:1}) associated with the bound states of the $(A-1)$ system can be
expressed in terms of the ground state one-body density matrix of the $A$
nucleon system. In the case of a target nucleus with $J^{\pi }=0^{+}$, the
lowest $(n_{0}lj)$ bound state overlap function is determined by the
asymptotic behavior $(a\rightarrow \infty )$ of the corresponding partial
radial contribution ${\rho _{lj}(r,r}^{\prime }{)}$ of the one-body density
matrix:
\begin{equation}
\phi _{n_{0}lj}(r)={\frac{{\rho _{lj}(r,a)}}{{C_{n_{0}lj}~\exp
(-k_{n_{0}lj}\,a})/a}}~,  \label{eq:5}
\end{equation}
where the constants ${C_{n_{0}lj}}$ and ${k_{n_{0}lj}}$ are completely
determined by ${\rho _{lj}(r,r}^{\prime }{)}$. In this way, both $\phi
_{n_{0}lj}(r)$ and ${k_{n_{0}lj}}$ define the separation energy
\begin{equation}
\epsilon _{n_{0}lj}\equiv E_{n_{0}lj}^{(A-1)}~-~E_{0}^{(A)}=\frac{\hbar
^{2}~k_{n_{0}lj}^{2}}{2m}~  \label{eq:6}
\end{equation}
and the spectroscopic factor $S_{n_{0}lj}=\langle \phi _{n_{0}lj}\mid \phi
_{n_{0}lj}\rangle $. The procedure also yields the next bound state overlap
functions with the same multipolarity if they exist. The applicability
of this procedure has been demonstrated in Refs.\cite{Vn96,Vn97,Sto96}.

Thus having the procedure for estimating such important quantities as
spectroscopic factors and overlap functions one has simply to apply it to
some realistic one--body density matrices emerging from the CBF and Green
function methods. The latter are briefly discussed in this Section.

\subsection{The CBF theory}

The CBF theory starts from a trial many--particle wave function
\begin{equation}
\Psi (x_{1},...,x_{A})={\cal S}\left[ \prod_{i<j=1}^{A}\hat{F}%
(x_{i},x_{j})\right] \Phi (x_{1},...,x_{A}),  \label{eq:7}
\end{equation}
where $A$ is the number of the nucleons with particle coordinates $%
x_{1},x_{2},...,x_{A}$ which contain spatial, spin, and isospin variables, $%
{\cal S}$ is a symmetrization operator, and $\Phi $ is an uncorrelated
(Slater determinant) wave function normalized to unity and describing a
closed-shell spherical system. The correlation factor $\hat{F}$ is generally
written as
\begin{equation}
\hat{F}(x_{i},x_{j})=\sum_{n}h_{n}(|{\bf r}_{i}-{\bf r}_{j}|)\hat{O}_{ij}^{n}
\label{eq:8}
\end{equation}
with basic two--nucleon operators $\hat{O}^{n}$ inducing central,
spin--spin, tensor and spin--orbit correlations, either with or without
isospin exchange:
\begin{equation}
\begin{array}{lcl}
O_{ij}^{n=1,...,8} & = & 1,(\mbox{\boldmath $\tau$}_{i}.%
\mbox{\boldmath
$\tau$}_{j}),(\mbox{\boldmath $\sigma$}_{i}.\mbox{\boldmath $\sigma$}_{j}),(%
\mbox{\boldmath $\sigma$}_{i}.\mbox{\boldmath $\sigma$}_{j})(%
\mbox{\boldmath
$\tau$}_{i}.\mbox{\boldmath $\tau$}_{j}),S_{ij}, \\
&  & S_{ij}(\mbox{\boldmath $\tau$}_{i}.\mbox{\boldmath $\tau$}_{j}),{\bf L}.%
{\bf S},{\bf L}.{\bf S}(\mbox{\boldmath $\tau$}_{i}.\mbox{\boldmath
$\tau$}_{j}).
\end{array}
\label{eq:9}
\end{equation}
The corresponding one-body density matrix
\begin{equation}
N(x_{1},x_{1}^{\prime })=\frac{\langle \Psi |c^{\dagger }(x_{1}^{\prime
})c(x_{1})|\Psi \rangle }{\langle \Psi |\Psi \rangle }  \label{eq:10}
\end{equation}
has been calculated in \cite{Pi92} using the Monte Carlo techniques. In many
cases the low--order approximation (LOA) \cite{Ga71,Da82,Fl84,Be86} is used
for the ODM which consists in expanding the corresponding quantities up to
the first order of the function $\hat{h}(x_{i},x_{j};x_{i}^{\prime
},x_{j}^{\prime })=\hat{F} (x_{i}^{\prime },x_{j}^{\prime })\hat{F}%
(x_{i},x_{j})-1$. The LOA expression for the ODM is:
\begin{equation}
N(x_{1},x_{1}^{\prime })=N_{0}(x_{1},x_{1}^{\prime
})+N_{1}(x_{1},x_{1}^{\prime })+N_{2}(x_{1},x_{1}^{\prime }),  \label{eq:11}
\end{equation}
with
\begin{equation}
N_{0}(x_{1},x_{1}^{\prime })=\rho (x_{1},x_{1}^{\prime }),  \label{eq:12}
\end{equation}
\begin{equation}
N_{1}(x_{1},x_{1}^{\prime })=\int dx_{2}\hat{h}(x_{1},x_{2};x_{1}^{\prime
},x_{2})[\rho (x_{1},x_{1}^{\prime })\rho (x_{2},x_{2})-\rho
(x_{1},x_{2})\rho (x_{2},x_{1}^{\prime })],  \label{eq:13}
\end{equation}
\begin{equation}
N_{2}(x_{1},x_{1}^{\prime })=\int dx_{2}dx_{3}\hat{h}%
(x_{2},x_{3};x_{2},x_{3})\rho (x_{1},x_{2})[\rho (x_{2},x_{1}^{\prime })\rho
(x_{3},x_{3})-\rho (x_{2},x_{3})\rho (x_{3},x_{1}^{\prime })].
\label{1eq:14}
\end{equation}
The zeroth order contribution (\ref{eq:12}) is the uncorrelated ODM
associated with the Slater determinant $\Phi $. An important feature of the power
series cluster expansion is that sum rule properties like the normalization property
is fulfilled at any order of the expansion \cite{Faf98}. Thus, in our case the
conservation of the number of particles, i.e.,
\begin{equation}
\int dx_{1}N(x_{1},x_{1})=\int dx_{1}\rho (x_{1},x_{1})=A.  \label{eq:15}
\end{equation}
is ensured.

The overlap functions for $^{16}O$ have explicitly been constructed on the
basis of the ODM generated by a CBF--type correlated wave function in \cite
{Vn97}. The s.p. orbitals entering the Slater determinant $\Phi $ were taken
from a Hartree--Fock calculation with the Skyrme--III effective force. The
correlation factor $\hat{F}(x_{1},x_{2})$ obtained in \cite{Pi92} by
variational calculations with Argonne NN forces has been used. The
two--nucleon correlation factors were restricted to the central,
spin--isospin and tensor--isospin operators. Such a description allows one
to distinguish between the effects of different types of correlations on
quantities such as overlap functions and spectroscopic factors of quasihole
states.

\subsection{FHNC formalism within the CBF theory}

The CBF theory and the Fermi hypernetted chain technique have been extended
in \cite{Sa96} to study medium--heavy doubly--closed shell nuclei in the $jj$
coupling scheme with different single--particle wave functions for protons
and neutrons using isospin--dependent two--body correlations. These are the
first microscopic calculations for nuclei beyond $^{40}Ca$ which are a
necessary step towards a correct description of heavy nuclear systems based
on realistic nuclear Hamiltonian. The FHNC equations can be written in terms
of the one--body densities and the two--body distribution functions:
\begin{equation}
\rho_{1}^{\alpha }({\bf r})=\langle \Psi^{*} \sum_{k=1}^{A} \delta ({\bf r}-%
{\bf r}_{k}) P_{k}^{\alpha }\Psi \rangle,  \label{eq:16}
\end{equation}
\begin{equation}
\rho_{2,q}^{\alpha \beta}=\langle \Psi^{*}\sum_{k\neq l=1}^{A} \delta ({\bf r%
}-{\bf r}_{k}) P_{k}^{\alpha } \delta ({\bf r^{\prime }}-{\bf r}_{l})
P_{l}^{\beta } O_{kl}^{q}\Psi \rangle, \;\;\; q=1,...,4,  \label{eq:17}
\end{equation}
where $P_{k}^{\alpha }$ is the projection operator on the $\alpha=p,n$ state
of the $k$-nucleon. The index $q$ labels the operational component of $\rho
_{2,q}^{\alpha \beta }$ with $O_{12}^{q}$ characterizing the first four
channels from Eq.(\ref{eq:9}). The calculations in \cite{Sa96} have been
performed using central nucleon--nucleon interactions ($v_{4}$) with spin
and isospin dependence but without tensor and spin--orbit components:
\begin{equation}
v_{4}(1,2)=\sum_{q=1,4} v^{(q)}(r_{12})O_{12}^{q}.  \label{eq:18}
\end{equation}
The structure of the FHNC equations depends on the adopted correlation
function $f$. It has been shown in \cite{Sa96} that the isospin dependence of
the correlation function within this approximation scheme is weak.
This is due to the fact that within this approach the correlations mainly result
from the central short--range components of the NN interaction.
Therefore, in our
calculations we use ODM for $^{16}O$ obtained up to the first order in the
cluster expansion by adopting the Average Correlation Approximation (ACA)
\cite{Sa96}. It consists in using of unique correlation, independent on the
isospin of the nucleons. The ACA correlations are well reproduced by a sum
of two gaussians:
\begin{equation}
f(r)=1- \alpha_{1} e^{-\beta_{1}r^{2}}+\alpha_{2}e^{-\beta_{2}(r-x)^{2}},
\label{eq:19}
\end{equation}
with the parameters: $\alpha_{1}$=0.64, $\beta_{1}$=1.54 $fm^{-2}$, $%
\alpha_{2}$=0.11, $\beta_{2}$=3.51 $fm^{-2}$ and $x$=1.0 $fm$. They are
taken as variational parameters fixed by minimizing the FHNC ground--state
energy \cite{Sa96}. The second ingredient of these calculations is the set
of s.p. wave functions which have been generated by a mean--field potential
of Woods--Saxon type.

The same s.p. wave functions but a state--dependent correlation function
taken from nuclear matter FHNC calculations have been used in \cite{Sa97} to
construct the ODM. The effects of state--dependent correlations on nucleon
density and momentum distributions of various nuclei have been studied in
\cite{Sa97}. It has been shown that the correlation functions used in these
calculations lead to a general lowering of the density distributions in the
interior region and to high--momentum components of the momentum
distribution.

\subsection{The Green function approach}

Recent microscopic calculations of the one--body Green function for $%
^{16}O $ have demonstrated \cite{Mu95,Mut95} that the nucleon--nucleon
correlations induced by the short--range and tensor components of a
realistic interaction yield an enhancement of the momentum distribution at
high momenta. This enhancement originates from the spectral function at large
negative energies
and therefore should be observed in nucleon knockout reactions with large
energy transfer leaving the final nucleus at an excitation energy of about $%
100$ $MeV$. For a nucleus like $^{16}O$ with $J=0$ ground state angular
momentum the one--body density matrix can easily be separated into
sub--matrices of a given
orbital angular momentum $l$ and total angular momentum $j$. Within the
Green function approach \cite{Di92} the ODM in momentum representation can
be evaluated from the imaginary part of the single--particle Green function
by integrating
\begin{equation}
\rho _{lj}(k_{1},k_{2})=\int_{-\infty }^{\varepsilon _{F}}dE\frac{1}{\pi }%
Im(g_{lj}(k_{1},k_{2};E)),  \label{eq:20}
\end{equation}
where the energy variable $E$ corresponds to the energy difference between
the ground state of the $A$ particle system and the energies of the states
in the $(A-1)$--particle system (negative $E$ with large absolute value
correspond to high excitation energies of the residual system) and $%
\varepsilon _{F}$ is the Fermi energy. The single--particle Green function $%
g_{lj}$ (or the propagator) is obtained from the solution of the Dyson
equation
\begin{equation}
g_{lj}(k_{1},k_{2};E)=g_{lj}^{(0)}(k_{1},k_{2};E)+\int dk_{3}\int
dk_{4}g_{lj}^{(0)}(k_{1},k_{3};E)\Delta \Sigma
_{lj}(k_{3},k_{4};E)g_{lj}(k_{4},k_{2};E),  \label{eq:21}
\end{equation}
where $g^{(0)}$ refers to the Hartree--Fock propagator and $\Delta \Sigma
_{lj}$ represents contributions to the real and imaginary part of the
irreducible self--energy, which go beyond the Hartree--Fock approximation of
the nucleon self--energy used to derive $g^{(0)}$.

The results for the ODM have been analyzed in \cite{Po96} in terms of the
natural orbitals $\varphi_{\alpha }$ and the occupation numbers $n_{\alpha }$
in $^{16}O$ nucleus. Within the natural orbital representation they can be
determined by diagonalizing the one--body density matrix of the correlated
system. In this representation the radial ODM for each $lj$ subspace has the
form:
\begin{equation}
\rho_{lj}(r,r^{\prime })=\sum_{\alpha } n_{\alpha lj}\varphi_{\alpha
lj}^{*}(r) \varphi_{\alpha lj}(r^{\prime }).  \label{eq:22}
\end{equation}
The numerical results from \cite{Po96} show that the ODM can be described quite
accurately in
terms of four natural orbitals $(\alpha=1,...,4)$ for each partial wave $lj$
in the sum (\ref{eq:22}). The ODM generated in this way is used in our
calculations to construct the single--particle overlap functions.

\section{Numerical results}

The ODM obtained with the different methods mentioned in Section 2 have been
applied to
calculate overlap functions related to the $1s$ and $1p$ states in the $%
^{16}O$ nucleus. The ODM from \cite{Vn97,Po96} account for non-central
correlation effects. Therefore one obtains in these approaches different
results for $p_{3/2}$ and $p_{1/2}$ quasihole states. This allows
us to calculate the corresponding differential cross--sections of $%
^{16}O(p,d)^{15}O$ reactions leading to the ground $1/2^{-}$ state and $%
3/2^{-}$ excited states of the residual $^{15}O$ nucleus.

\subsection{Overlap functions and spectroscopic factors}

In this subsection we present the results for the overlap functions, the
spectroscopic factors and the neutron separation energies calculated using
the procedure of Eqs.(\ref{eq:1})-(\ref{eq:6}) (see also \cite{Vn93}).

The resulting overlap functions are compared with the HF wave function in
Fig.~1. The HF wave function has been calculated in a self-consistent way
using the Skyrme-III interaction. It is the uncorrelated basis function,
which has also been used in \cite{Vn97}.
It can be seen from Fig.~1 that the overlap functions and the HF
wave function are rather similar. This is not only true for the example of $1s$
states exhibited in this figure but also for the $1p$ hole states in $^{16}O
$. This justifies the use of shell--model orbitals instead of overlap
functions in calculating the nucleon knock-out cross section using the
Plane Wave Impulse Approximation
for such kind of nuclear states. The changes in the shape from
the original mean--field wave functions are rather small and might be absorbed
by a suitable readjustment of the parameters of the single--particle potential
used to determine the corresponding wave functions. The
inclusion of short-range as well as tensor correlations leads to an
enhancement of the values of the
corresponding overlap functions in the interior region and a depletion in
the tail region in the coordinate space. As expected, in the momentum space
these effects lead to a shift of the overlap functions from the
low-- to the high--momentum region in comparison with the mean--field wave
functions as it is shown in Fig.~2. This fact is important because the
squared overlap functions in the momentum space determine the
single--particle momentum distribution representing the transition to a
given single--particle state of the residual nucleus. This single--particle
momentum distribution can be obtained experimentally, e.g. from $%
(e,e^{\prime }p)$ reactions, by integrating the data for the spectral
function over the energy interval which includes the peak of the transition.

The values of the spectroscopic factors and the separation energies deduced
from the calculations with different ODM are listed in Table 1. It is seen
that the separation energies derived from ODM are in acceptable agreement with
the corresponding single-particle energies obtained in
self--consistent Hartree--Fock calculations.
As it has been shown already in \cite{Vn97}, the use of
single-particle wave functions which have realistic exponential
asymptotics leads to
separation energies of the quasihole states which are close to the original
mean--field single-particle energies.

The calculated spectroscopic factors, however,
differ significantly from the mean--field value.  The
nucleon--nucleon correlations lead to a depletion of the states which are
below the Fermi level of the independent particle approach.
The spectroscopic factors of the $s$ and $p$ states in $%
^{16}O $ obtained within the JCM (0.94 and 0.953, respectively) \cite{Sto96}
are somewhat smaller than the values obtained from the calculations which
include only central channel of the interaction (about 0.98) \cite{Vn97}.
Although both central correlation functions have a comparable range,
the first one is more effective at small $r$, going to zero for $r=0$. Therefore
the correlation effects induced by the central correlation function in this
approach are stronger leading to a smaller spectroscopic factor.
The same holds for the ODM generated in \cite{Sa96,Sa97}. The
comparison of the spectroscopic factors also shows that the tensor correlations,
which are taken into account in \cite{Vn97,Po96}, are responsible for a large
part of the depletion of the
occupied states. The central correlations generate a depletion of 1-2 $\%$
only, whereas the inclusion of the tensor channel leads to a
depletion of 7-11 $\%$ \cite{Vn97}. The spectroscopic factors for the $%
p_{3/2}$ and $p_{1/2}$ quasihole states in $^{16}O$ found in \cite{Vn97} are
about 0.90--0.91. Our calculations based on the Green function theory
yield similar results which indicates that about 10 $\%$ of the $1p$%
--strength is removed by the short--range and tensor correlations. Here one
should keep in mind that an additional depletion or reduction of the
spectroscopic factors, which is not included in the approaches presented here,
is due to long-range correlations\cite{skour}.

In Table 1 the spectroscopic factors $S$ are given together with the natural
occupation numbers $N$ \cite{An93}  calculated after diagonalizing the
corresponding one--body density matrices. The comparison shows that the
results satisfy the general property $S_{nlj}\leq N_{nlj}^{max}$, i.e. in
each $lj$ subspace the spectroscopic factor $S_{nlj}$ is smaller than the
largest natural occupation number $N_{nlj}^{max}$ \cite{Vn93}. The trend of
the calculated spectroscopic factors follows that of the natural occupation
numbers.

In the case of the Green function approach \cite{Po96,Mu95} one can compare the
spectroscopic factor for the overlap function and the separation energy
derived from the ODM with the occupation probability and the energy of the
corresponding quasihole state listed in Table 2 of \cite{Mu95}. The occupation
probabilities for the quasihole states (0.780, 0.898, 0.914 for $s_{1/2}$,
$p_{1/2}$ and $p_{3/2}$, respectively) are slightly smaller than the
spectroscopic factors listed in Table 1, indicating that the continuum contribution
to the spectral function is non-negligible. The difference is largest for the $s_{1/2}$ state. The
absolute energy of the quasihole state is slightly larger for the $s_{1/2}$
(34.3 MeV) than the corresponding separation energy (31.12 MeV) deduced from the
ODM. In the case of the $p$ states the quasihole energies are smaller (14.14 MeV
and 17.9 MeV) than the separation energies.

\subsection{Differential cross--sections}

In order to explore whether an analysis of experimental data is sensitive to
the differences in the overlap functions derived from the various many-body
theories, we are now going to employ a very simple model for calculating
cross--sections of $^{16}O(p,d)$
reactions. The differential cross--section for such pick--up
processes can be written in the form:
\begin{equation}
\frac{d\sigma _{pd}^{lsj}(\theta )}{d\Omega }=\frac{3}{2}\frac{S_{lsj}}{2j+1}%
\frac{D_{0}^{2}}{10^{4}}\sigma _{DW}^{lsj}(\theta ),  \label{eq:23}
\end{equation}
where $S_{lsj}$ is the spectroscopic amplitude, $j$ is the total angular
momentum of the final state, $D_{0}^{2}\approx 1.5\times 10^{4}$ $MeV.fm^{3}$
is the $p-n$ interaction strength in the zero--range approximation
and $\sigma_{DW}^{lsj}$ is the cross--section calculated by the DWUCK4 code
\cite{La93}.
For our purposes the standard Distorted Wave Born Approximation (DWBA)
form factor has been replaced by the s.p. overlap function derived from
the one--body density matrix calculations. In this case no extra
spectroscopic factor $S_{lsj}$ in eq.(\ref{eq:23}) is needed,
since our overlap functions already include the associated
spectroscopic factors. The results for the differential cross--sections for
the transitions to the ground $1/2^{-}$ state and to the excited $3/2^{-}$
state in $^{15}O$ nucleus at different incident proton energies $E_{p}$%
=31.82, 45.34 and 65 $MeV$  are given in Figures 3--6. A comparison with the
experimental data from \cite{Pre70,Ro75} is also made. The optical
potentials parameter values have been taken in each case to be the same as
in the corresponding standard DWBA calculations.

As can be seen in Figs.3--5 the use of all overlap functions for the
transition to the ground $1/2^{-}$ state leads to a qualitative agreement
with the experimental data reproducing the amplitude of the first maximum
and qualitatively the shape of the differential cross--section. The differences
between the results obtained from the various approaches are small but visible.

We emphasize that our results are without any additional normalization while
the standard DWBA curves need multiplication by the fitting parameter, i.e.,
the spectroscopic factor. The values of the spectroscopic factors
obtained from the standard DWBA procedure are given in Table 2 and
can be compared with our spectroscopic factors from Table 1. It is seen that
the value of the DWBA spectroscopic factors for the $1/2^{-}$ state exceeds the
maximum allowed value of unity.

\section{Conclusions}

The results of the present work can be summarized as follows: \newline
i) Single--particle overlap functions, spectroscopic factors and separation
energies are calculated from the one--body density matrices, which were
derived using different approximations to determine the correlated wave function
for the ground state of $^{16}O$. \newline
ii) The overlap functions extracted from ODM calculated within the CBF and
Green function theories are peaked at smaller distance in the interior
region of the nucleus compared with  Hartree-Fock wave functions. \newline
iii) Considering the role of the central and the tensor correlations it is
found that the correlation effects on the spectroscopic factors of the hole
states are dominated by the tensor channel of the interaction. \newline
iv) The absolute values of the differential cross--sections of $^{16}O(p,d)$
pick--up reaction at various incident energies are calculated by using the
 overlap functions. The resulting angular distributions are in a
qualitative agreement with the experimental cross--sections of the
transitions to the ground $1/2^{-}$ and excited $3/2^{-}$ states of the
residual nucleus $^{15}O$.

\acknowledgments

The authors are grateful to Dr. G. Co' for providing us the results for ODM
from \cite{Sa96,Sa97} and to Dr. C. Giusti for the valuable discussions.
This work was partly supported by the Bulgarian National Science Foundation
under the Contracts Nrs.$\Phi $--527 and $\Phi $--809.

\newpage
\noindent
\noindent {\bf Table 1:} Spectroscopic factors ($S$),
occupation numbers ($N$) and separation energies ($\varepsilon $) deduced
from the calculations with different ODM for $^{16}O$ \vspace{1cm}

\begin{center}
\begin{tabular}{cccccccl}
\hline\hline
&  & \multicolumn{2}{c}{$1s$} &  &  & \multicolumn{2}{c}{$1p$} \\
\cline{2-4}\cline{6-8}
ODM & $S$ & $N$ & $\varepsilon , MeV$ &  & $S$ & $N$ & $\varepsilon , MeV$
\\ \cline{2-4}\cline{6-8}\cline{1-4}\cline{6-8}
HF & 1.000 & 1.000 & 34.89 &  & 1.000 & 1.000 & 14.84 ($j$=1/2) \\
&  &  &  &  & 1.000 & 1.000 & 21.17 ($j$=3/2) \\
JCM\cite{Sto96} & 0.940 & 0.950 & 35.82 &  & 0.953 & 0.965 & 17.48 \\
CBF\cite{Vn97} & 0.883 & 0.884 & 37.70 &  & 0.912 & 0.929 & 16.79 ($j$=1/2)
\\
&  &  &  &  & 0.909 & 0.917 & 23.30 ($j$=3/2) \\
CBF\cite{Sa96} & 0.977 & 0.978 & \ 34.85 &  & 0.981 & 0.984 & 20.09 \\
CBF\cite{Sa97} & 0.980 & 0.980 & 34.80 &  & 0.983 & 0.985 & 20.05 \\
GFM\cite{Po96} & 0.904 & 0.933 & 31.12 &  & 0.905 & 0.916 & 16.68 ($j$=1/2)
\\
&  &  &  &  & 0.915 & 0.922 & 20.74 ($j$=3/2) \\ \hline\hline
\end{tabular}
\end{center}

\vspace{1cm}

{\bf Table 2:} Spectroscopic factors $S_{DWBA}/(2j+1)$ deduced from the
standard DWBA calculations for the $^{16}O(p,d)$ reactions leading to the $%
1/2^{-}$ ground and $3/2^{-}$ excited states of the $^{15}O$ nucleus \vspace{%
1cm}

\begin{center}
\begin{tabular}{cccc}
\hline\hline
Proton energy, $MeV$ & $1/2^{-}$ & $3/2^{-}$ &  \\ \cline{2-3}\cline{1-3}
$E_{p}=31.82$ & 1.05 & 0.85 &  \\
$E_{p}=45.34$ & 1.22 & 0.83 &  \\
$E_{p}=65$ & 1.62 & 0.54 &  \\ \hline\hline
\end{tabular}
\end{center}

\vfil
\begin{figure}
\epsfysize=8.0cm
\begin{center}
\makebox[16.0cm][c]{\epsfbox{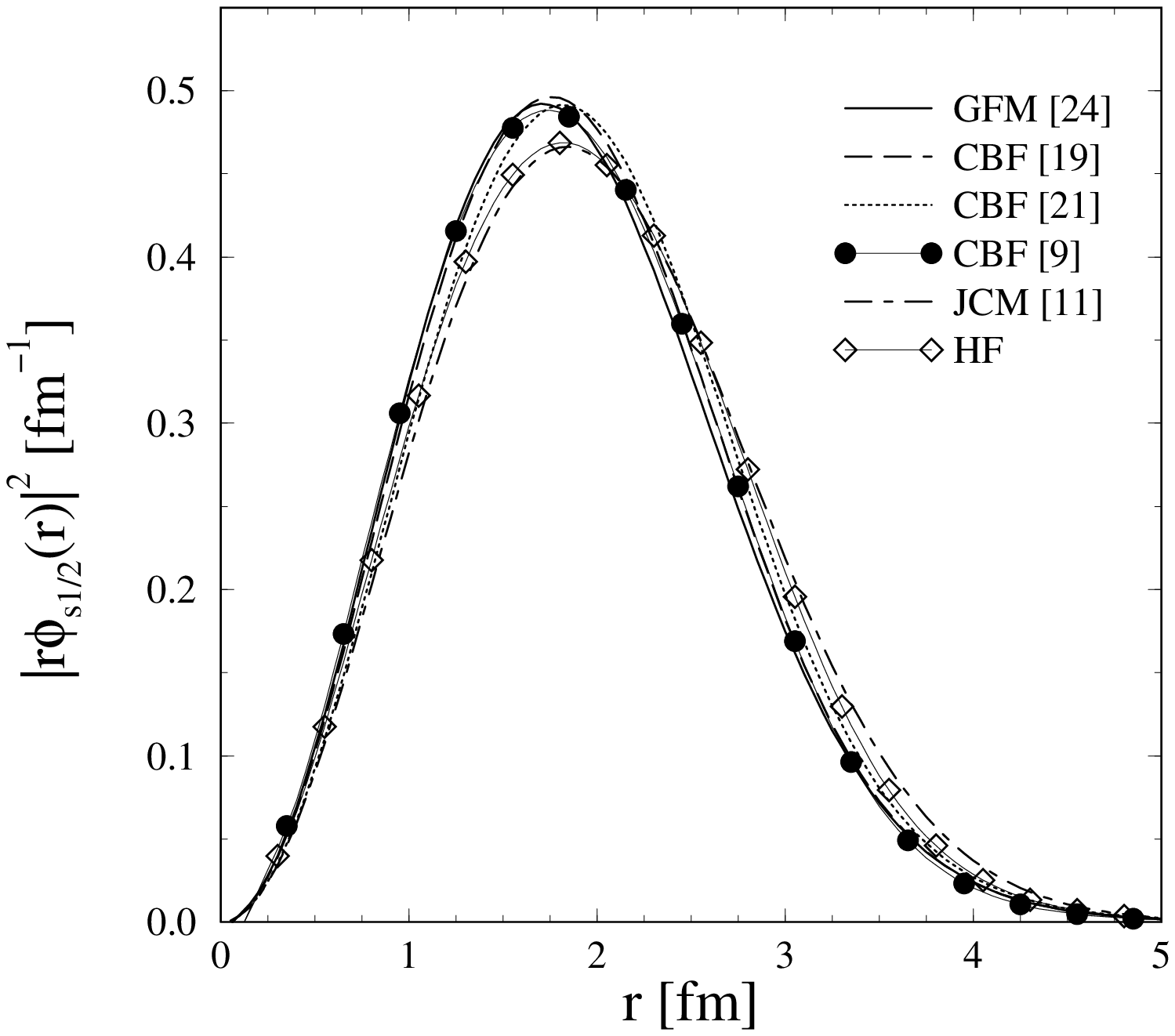}}
\end{center}
\caption{
Overlap functions ($r\phi(r)$ squared) for the
neutron $s_{1/2}$ quasihole state in $^{16}O$. All curves are normalized to
unity.}
\end{figure} 

\begin{figure}
\epsfysize=8.0cm
\begin{center}
\makebox[16.0cm][c]{\epsfbox{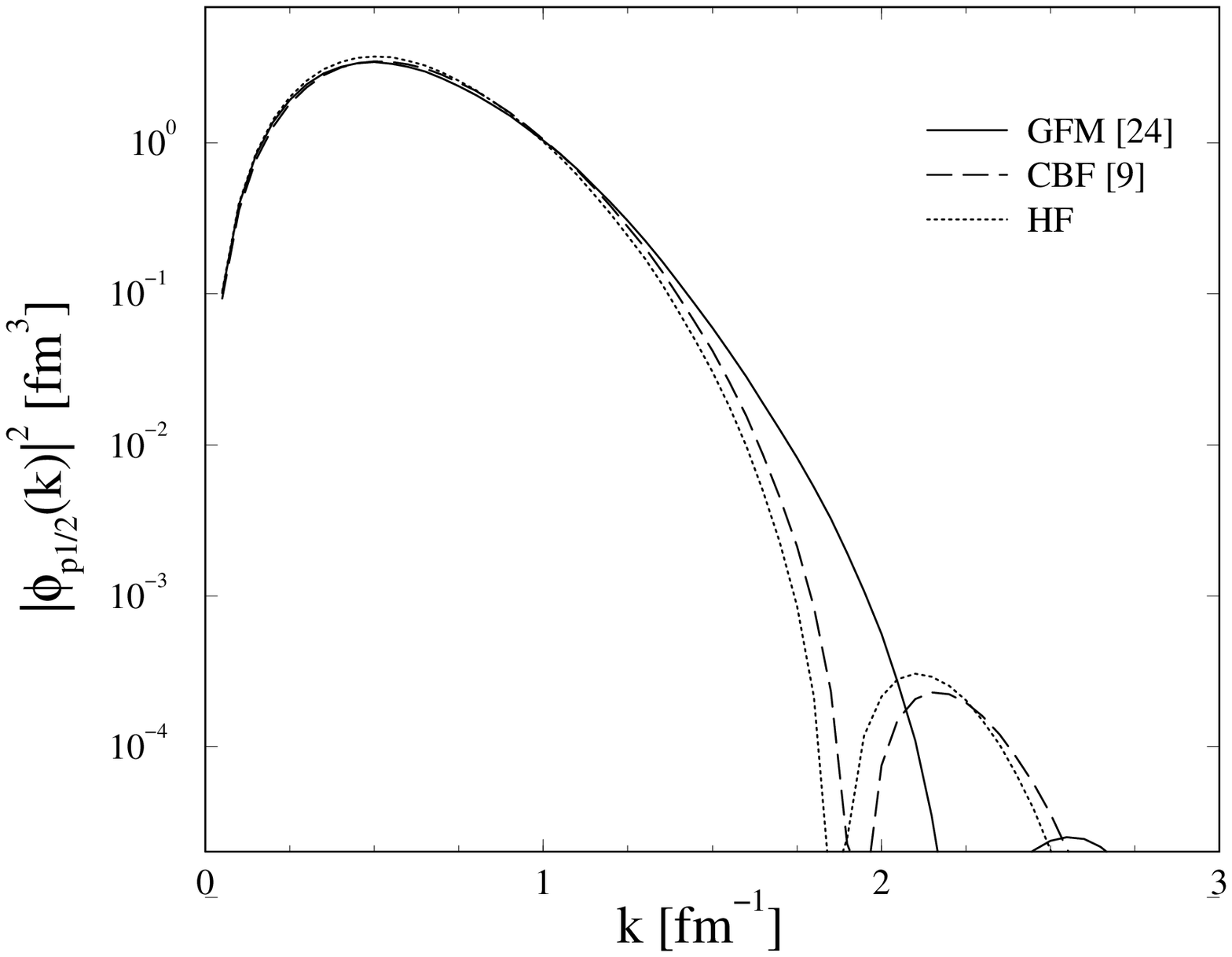}}
\end{center}
\caption{
Overlap functions ($\phi(k)$ squared) for the neutron $p_{1/2}$
quasihole state in $^{16}O$. All curves are normalized to unity.} 
\end{figure} 

\begin{figure}
\epsfysize=12.0cm
\begin{center}
\makebox[16.0cm][c]{\epsfbox{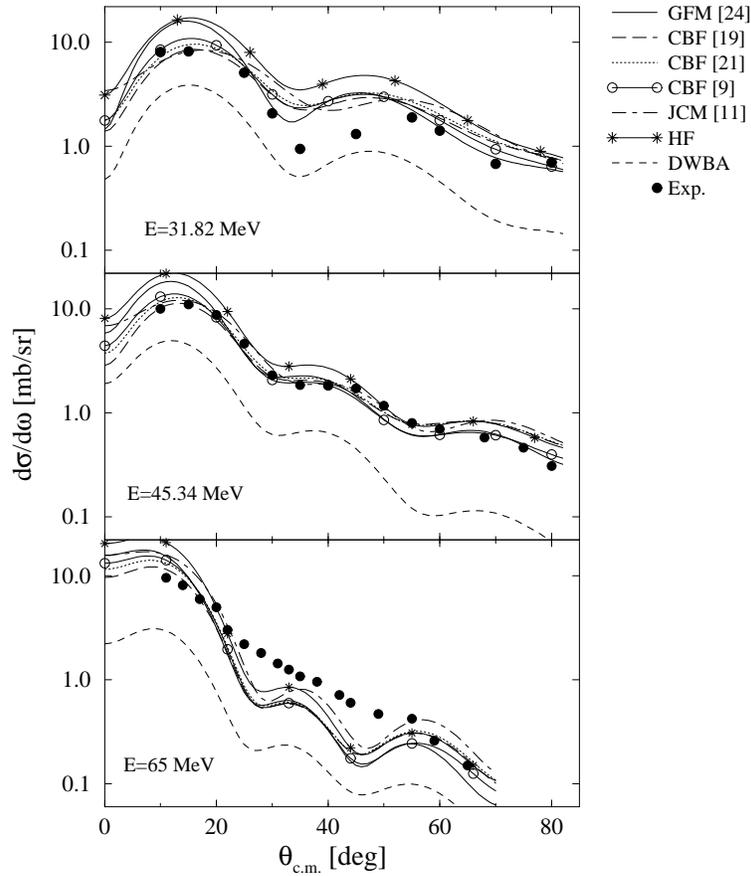}}
\end{center}
\caption{Differential cross--section for the $^{16}O(p,d)$ reaction at incident
proton energy $E_{p}=31.82$ $MeV$, $E_{p}=45.34$ $MeV$ and $E_{p}=65$ $MeV$ to
the $1/2^{-}$ ground state in $^{15}O$. The curves refering to
results which are derived from various ODM are labeled by the
reference, in which this ODM has been calculated. The
experimental data \protect\cite{Pre70} are given by the full circles.}
\end{figure}

\begin{figure}
\epsfysize=8.0cm
\begin{center}
\makebox[16.0cm][c]{\epsfbox{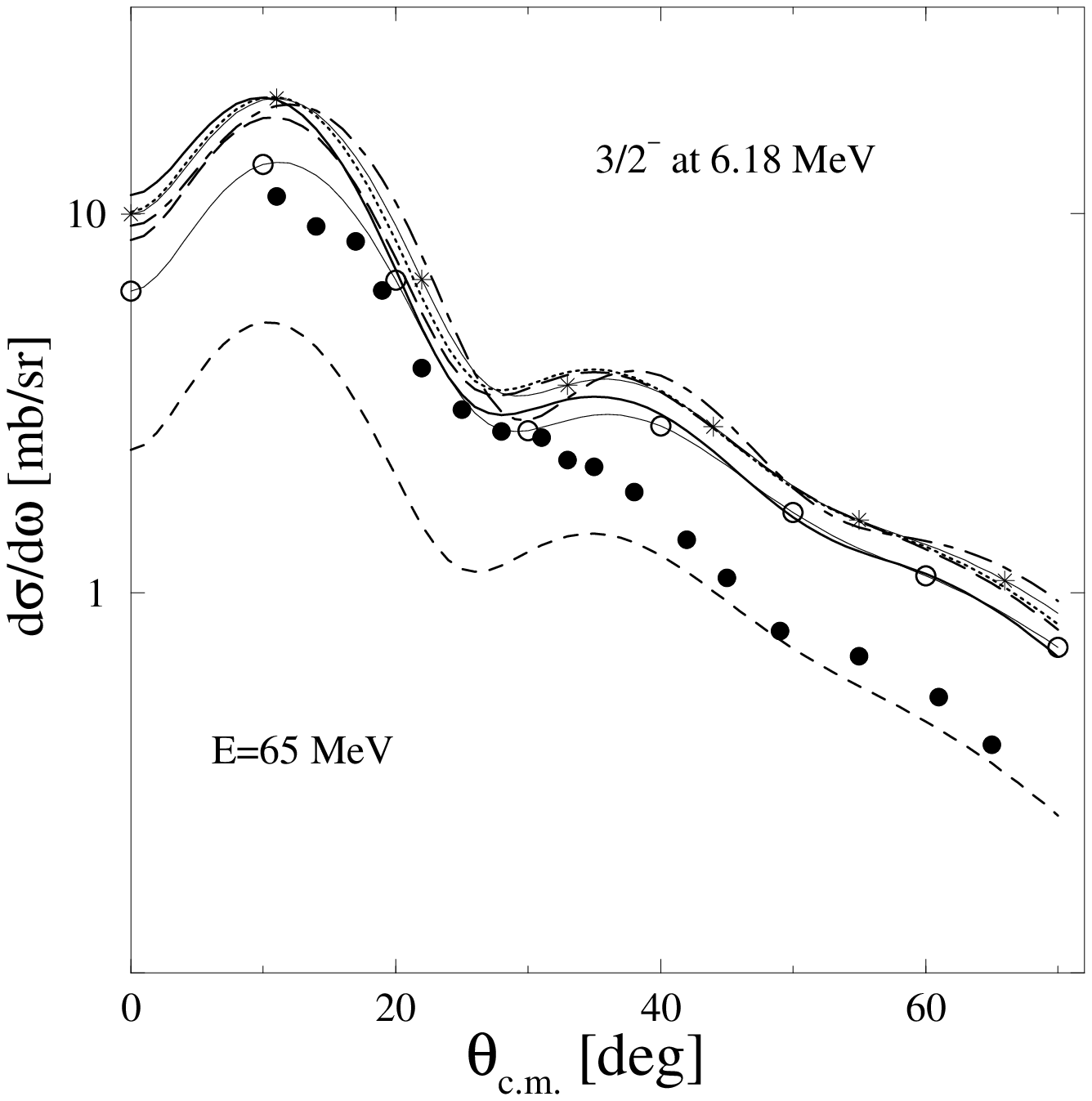}}
\end{center}
\caption{Differential cross--section for the $^{16}O(p,d)$ reaction at $%
E_{p}=65$ $MeV$ incident energy to the $3/2^{-}$ excited state in
$^{15}O$. The notations are the same as in Fig.3. The experimental data are
taken from \protect\cite{Ro75}.}
\end{figure}


\begin{references}
\bibitem{An93}  A.N. Antonov, P.E. Hodgson, and I.Zh. Petkov, {\it Nucleon
Correlations in Nuclei} (Springer-Verlag, Berlin, 1993).

\bibitem{Lap93}  L. Lapik\'{a}s, Nucl. Phys. {\bf A553}, 297c (1993).

\bibitem{Bo94}  I. Bobeldijk et al., Phys. Rev. Lett. {\bf 73}, 2684 (1994).

\bibitem{Bl95}  K.I. Blomqvist et al., Phys. Lett. B {\bf 344}, 85 (1995).

\bibitem{Vn93}  D. Van Neck, M. Waroquier, and K. Heyde, Phys. Lett. B {\bf %
314}, 255 (1993).

\bibitem{An95}  A.N. Antonov, M.V. Stoitsov, M.K. Gaidarov, S.S. Dimitrova,
and P.E. Hodgson, J. Phys. {\bf G21}, 1333 (1995).

\bibitem{Be65}  T. Berggren, Nucl. Phys. {\bf 72}, 337 (1965).

\bibitem{Vn96}  D. Van Neck, A.E.L. Dieperink, and M. Waroquier, Z. Phys.
{\bf A355}, 107 (1996); Phys. Rev. C {\bf 53}, 2231 (1996).

\bibitem{Vn97}  D. Van Neck, L. Van Daele, Y. Dewulf, and M. Waroquier,
Phys. Rev. C {\bf 56}, 1398 (1997).

\bibitem{Vn98}  D. Van Neck, M. Waroquier, A.E.L. Dieperink, S.C. Pieper,
and V.R. Pandharipande, Phys. Rev. C {\bf 57}, 2308 (1998).

\bibitem{Sto96}  M.V. Stoitsov, S.S. Dimitrova, and A.N. Antonov, Phys. Rev.
C {\bf 53}, 1254 (1996).

\bibitem{Ge96}  W.J.W. Geurts, K. Allaart, W.H. Dickhoff, and H. M\"{u}ther,
Phys. Rev. C {\bf 53}, 2207 (1996).

\bibitem{An88}  A.N. Antonov, P.E. Hodgson, and I.Zh. Petkov, {\it Nucleon
Momentum and Density Distributions} (Clarendon Press, Oxford, 1988).

\bibitem{Ma91}  C. Mahaux and R. Sartor, Adv. Nucl. Phys. {\bf 20}, 1 (1991).

\bibitem{Pi92}  S.C. Pieper, R.B. Wiringa, and V.R. Pandharipande, Phys.
Rev. C {\bf 46}, 1741 (1992).

\bibitem{Co92}  G. Co', A. Fabrocini, S. Fantoni, and I.E. Lagaris, Nucl.
Phys. {\bf A549}, 439 (1992).

\bibitem{De83}  F. Dellagiacoma, G. Orlandini, and M. Traini, Nucl. Phys.
{\bf A393}, 95 (1983).

\bibitem{Ja55}  R. Jastrow, Phys. Rev. {\bf 98}, 1479 (1955).

\bibitem{Sa96}  F. Arias de Saavedra, G. Co', A. Fabrocini, and S. Fantoni,
Nucl. Phys. {\bf A605}, 359 (1996).

\bibitem{Fa98}  A. Fabrocini, F. Arias de Saavedra, G. Co', and P.
Folgarait, Phys. Rev. C {\bf 57}, 1668 (1998).

\bibitem{Sa97}  F. Arias de Saavedra, G. Co', and M.M. Renis, Phys. Rev. C
{\bf 55}, 673 (1997).

\bibitem{Am98}  J.E. Amaro, A.M. Lallena, G. Co', and A. Fabrocini, Phys.
Rev. C {\bf 57}, 3473 (1998).

\bibitem{Di92}  W.H. Dickhoff and H. M\"{u}ther, Rep. Prog. Phys. {\bf 55},
1947 (1992).

\bibitem{Po96}  A. Polls, H. M\"{u}ther and W.H. Dickhoff, {\it Proceedings
of Conference on Perspectives in Nuclear Physics at Intermediate Energies},
Trieste, 1995, edited by S. Boffi, C. Ciofi degli Atti, and M.M. Giannini,
p.308 (World Scientific, Singapore, 1996).

\bibitem{Mu95}  H. M\"{u}ther, A. Polls, and W.H. Dickhoff, Phys. Rev. C
{\bf 51}, 3040 (1995).

\bibitem{Mut95}  H. M\"{u}ther, G. Knehr, and A. Polls, Phys. Rev. C {\bf 52}%
, 2955 (1995).

\bibitem{Ci96}  C. Ciofi degli Atti and S. Simula, Phys. Rev. C {\bf 53},
1689 (1996).

\bibitem{Mu94}  H. M\"{u}ther and W.H. Dickhoff, Phys. Rev. C {\bf 49}, R17
(1994).

\bibitem{Sto93}  M.V. Stoitsov, A.N. Antonov, and S.S. Dimitrova, Phys. Rev.
C {\bf 47}, R455 (1993); Phys. Rev. C {\bf 48}, 74 (1993); Z. Phys. {\bf A345%
}, 359 (1993).

\bibitem{Di97}  S.S. Dimitrova, M.K. Gaidarov, A.N. Antonov, M.V. Stoitsov,
P.E. Hodgson, V.K. Lukyanov, E.V. Zemlyanaya, and G.Z. Krumova, J. Phys.
{\bf G23}, 1685 (1997).

\bibitem{Ga71}  M. Gaudin, J. Gillespie, and G. Ripka, Nucl. Phys. {\bf A176}%
, 237 (1971).

\bibitem{Da82}  M. Dal R\`{i}, S. Stringari, and O. Bohigas, Nucl. Phys.
{\bf A376}, 81 (1982).

\bibitem{Fl84}  M.F. Flynn, J.W. Clark, R.M. Panoff, O. Bohigas, and S.
Stringari, Nucl. Phys. {\bf A427}, 253 (1984).

\bibitem{Be86}  O. Benhar, C. Ciofi degli Atti, S. Liuti, and G. Salm\'{e},
Phys. Lett. B {\bf 177}, 135 (1986).

\bibitem{Faf98}  S. Fantoni and A. Fabrocini, in {\it Lecture Notes in Physics},
Proceedings of the European Summer School on Microscopic Quantum Many Body
Theories and their Applications, Valencia, 1997, edited by J. Navarro and A.
Polls (Springer-Verlag, Berlin), 119 (1998).

\bibitem{skour}  K. Amir-Azimi-Nili, H. M\"uther, L.D. Skouras, and A.Polls,
Nucl. Phys. {\bf A604}, 245 (1996).

\bibitem{La93}  K. Langanke, J.A. Maruhn, and S.E. Koonin, {\it %
Computational Nucl. Phys. 2: Nucl. Reactions}, (Springer-Verlag,
Berlin-Heidelberg-New York), 88 (1993).

\bibitem{Pre70}  B.M. Preedom, J.L. Snelgrove, and E. Kashy, Phys. Rev. C
{\bf 1}, 1132 (1970).

\bibitem{Ro75}  P.G. Roos, S.M. Smith, V.K.C. Cheng, G. Tibell, A.A. Cowley,
and R.A. Riddle, Nucl. Phys. {\bf A255}, 187 (1975).
\end{references}
\end{document}